Comment on:

"Mysterious abrupt carbon-14 increase in coral contributed by a comet"
 Yi Liu et al.
 Scientific Reports 4, Article number: 3728  doi:10.1038/srep03728

Unfortunately, Liu et al. contains a number of errors and omissions which compromise its conclusions. These have to do with the amount of $^{14}$C which is necessary to deposit in the atmosphere in order to see a perturbation like that in 774 AD, and the ability of a comet to do so. They find in their coral data a $^{14}$C enhancement comparable to that in earlier work, but constrain it to take place on a much shorter timescale, of order two weeks. This is highly significant. They discuss the amount necessary to produce a 4.5% increase in atmospheric $^{14}$C. It must be assumed that corals, many of which are growing in shallow water, are able to rapidly absorb atmospheric gases. Since the exchange of atmospheric carbon with the ocean and surface life is rather slow, of order a decade, the relevant quantity is the ratio of new $^{14}$C to that resident in the atmosphere. Liu et al. give this number as 150 metric tonnes. However, this number is close to estimates[1] of the total quantity in the whole biosphere. The atmospheric mass of $^{14}$C is 500 to 850 kg[1]—leading to an error of a factor of 200 or so.  Taking 600 kg, we need 27 kg of $^{14}$C added to the atmosphere, much less than the many tonnes required by their estimate.

The other serious problem concerns their estimate of the $^{14}$C content of a comet ($10^{-7}$ by mass) which has the opposite effect on the computation.  We have computed the cosmogenic nuclide content of comets[2]. We find that observably large perturbations in the atmospheric $^{14}$C budget require large, long-period comets. Long-period comets are thought to originate outside the heliosphere, where they are exposed to increased cosmic rays, sufficient to induce the formation of $^{14}$C and other species.

More importantly, the cosmic ray showers which form these species rarely penetrate more than 20 meters into the body of the comet. The mass fraction of $^{14}$C is determined by a steady-state between creation and decay. So the amount of cosmogenic nuclei scales with the suface area, not the mass of the object. Meteorite data on $^{14}$C quoted by Liu et al. are irrelevant, because they are so small that their entire mass is exposed to the full cosmic ray flux. We find that the mass fraction of $^{14}$C for long-period comets runs from about $10^{-11}$ for a $10^7$ kg comet to about $10^{-14}$ for a $10^{16}$ kg comet. The coma is not particularly relevant here, as it forms when the comet approaches the Sun. So, to make the kind of perturbation described by Liu et al., the mass of the comet would have to lie between about $10^{14}$ kg and $3 \times 10^{15}$ kg. This can be compared with a Tunguska object mass estimated (with considerable uncertainty) to have a mass around $10^8$ kg and a hypothetical Younger Dryas impactor[3] at $5 \times 10^{13}$ kg. Certainly such an object would initiate continent-scale devastation, if not more, and could not have happened within historical times without major documentation.

A solar proton event is consistent with the data, and events that are improbable (except over long geological timescales) such as gamma-ray bursts are not required.

As this comment was being prepared, an eprint by Usoskin appeared (arXiv:1401.5945) which reached essentially the same conclusions.

Adrian Melott